# Structural, magnetic, and magnetocaloric properties of Fe$_2$CoAl Heusler nanoalloy

Aquil Ahmad*[a], Srimanta Mitra[b,c], S. K. Srivastava[b] and A. K. Das*[b]

[a]*Department of Physics, School of Electrical and Electronics Engineering, SASTRA Deemed University, Thanjavur-613401, Tamilnadu, India*

[b]*Department of Physics, Indian Institute of Technology Kharagpur, Kharagpur-721302, India*

[c]*Space Applications Center, ISRO, Ahmedabad-380015, India*

E-mails: aquilahmad@phy.sastra.edu  (A. Ahmad) and amal@phy.iitkgp.ernet.in (A. K. Das)

**\*** *Corresponding authors.*

**Abstract**

Spherical nanoparticles (NPs) of size 14±7 nm, made of intermetallic Fe$_2$CoAl (FCA) Heusler alloy, are synthesized via the co-precipitation and thermal deoxidization method. X-ray diffraction (XRD) and selected area electron diffraction (SAED) patterns confirm that the present nanoalloy is crystalized in A2-disordered cubic Heusler structure. Magnetic field (H) and temperature (T) dependent magnetization (M) results reveal that the NPs are soft ferromagnetic (FM) with high saturation magnetization (M$_s$) and Curie temperature (T$_c$). Fe$_2$CoAl nanoalloy do not follow the Slater Pauling (SP) rule, possibly because of the disorder present in the system. We also investigate its magnetic phase transition (MPT) and magnetocaloric (MC) properties. The peak value of $-\Delta S_M$ (entropy change) vs T curve at a magnetic field change of 20 kOe corresponds to about 2.65 J/kg-K, and the observed value of refrigeration capacity (RCP) is as large as 44 J/kg, suggesting a large heat conversion in magnetic refrigeration cycle. The Arrott plot and the nature of the universal curve accomplish that the FM to paramagnetic (PM) phase transition in Fe$_2$CoAl nanoalloy is of second-order. The present study suggests that the Fe$_2$CoAl nanoscale system is proficient, useful and a good candidate for the spintronics application and opens up a window for further research on full-Heusler based magnetic refrigerants.

## 1. Introduction

A significant enhancement over conventional electronics is possible with spintronics, where aside from charge of the electron, its spin also plays an important role in order to transfer and storage of the information [1]. The numerous Heusler compounds showing half-metallic character, low Gilbert damping together with the high $T_c$ and magnetic moment, have proven their potentiality for spintronics applications [2-5]. The scarce half-metallicity arises due to their unique electronic band structures at the fermi energy ($E_F$) of which one-spin (up or down) band behaves as a metal, while the other-spin (down or up) behaves as a semiconductor, and unveils an energy-gap at the $E_F$ [3, 5]. Hence, Heusler alloys (HAs) may increase the performance of spin-based devices relying upon magnetic tunnel junction and giant magnetoresistance or spin transfer torque [6]. Quite a few materials earlier have shown up to 100% spin polarization (SP) at the $E_F$ such as $CrO_2$ [7], $Fe_3O_4$ [8], EuS [9] and EuO [10], but most of them are neither appropriate for a high spin transport nor suitable with Si platforms because of poor electrical contacts [1, 2]. Subsequently, the hunting of novel materials for high SP is a consistently growing research field. In recent years, HAs have attracted enormous interest to the researchers due to their unique and multifunctional properties such as half-metallicity [11-14], magnetocaloric (MC) effect [15-17], thermoelectric effect [18] and catalytic behavior [19-22], shape memory effect [23], spin injection [24], and spin filtering [25]. To understand the fundamentals of MC effect, the measurement techniques for characterizing the samples and a comparative investigation of different MC materials along with probable routes to improve their performance, readers are referred to these articles [17, 26-32]. Notably, the physical properties of HAs can easily be tuned preserving their high magnetic

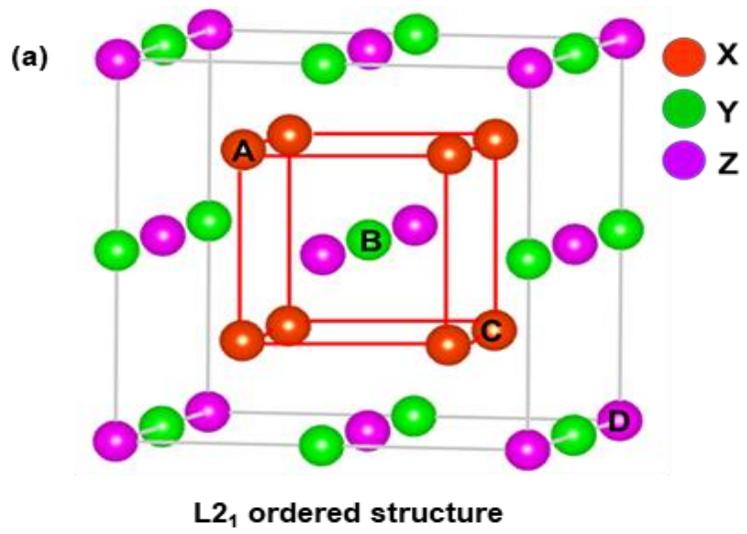

**L2$_1$ ordered structure**

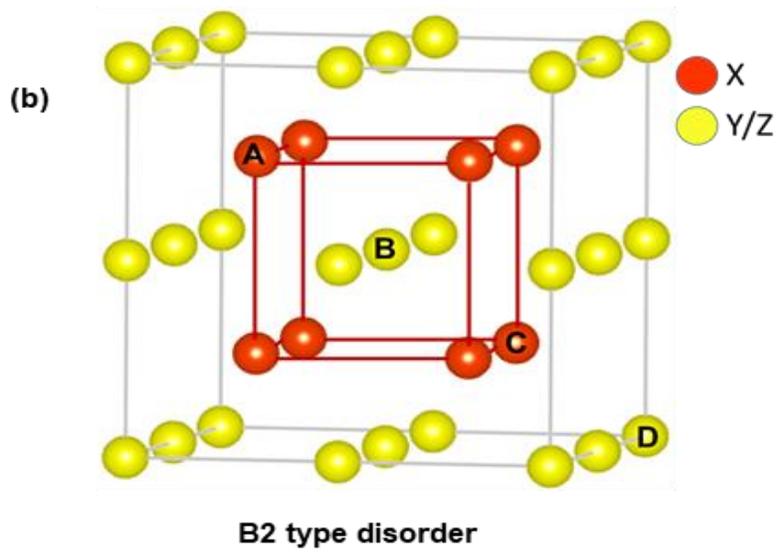

**B2 type disorder**

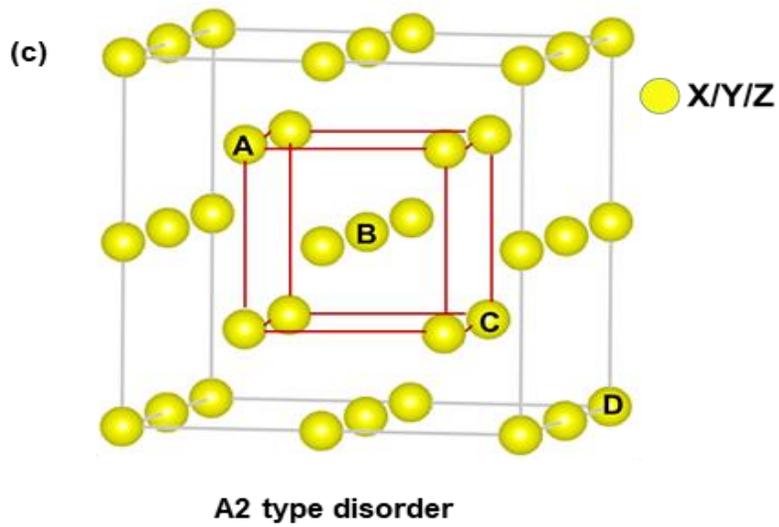

**A2 type disorder**

**Fig. 1.** Overview of most notably types of ordered/disordered crystallographic structures of $X_2YZ$ type full Heusler compounds: (a) $L2_1$ ordered structure, (b) B2- type disordered structure, and (c) A2 type disordered (or fully disordered) structure. The four possible Wyckoff sites: A, B, C, and D are also shown in the figures. Crystal structures are generated using VESTA software.

moments and Curie temperature, by altering their elemental compositions and/or partial substitution with other elements [33-36]. $X_2YZ$ type full Heusler compounds (X and Y: transition metals; Z: main group element) generally crystallized in highly ordered cubic $L2_1$ phase of $Cu_2MnAl$ prototype at the room temperature. The $L2_1$ phase belongs to the $Fm\bar{3}m$ space group (#225) [6, 14, 37] and can be understood as four interpenetrating fcc sublattices of which the X atoms occupy the 8c (0.25, 0.25, 0.25 & 0.75, 0.75, 0.75) Wyckoff site, Y atoms occupy the 4b (0.5, 0.5, 0.5) site, and lastly the Z atoms reside at the 4a (0, 0, 0) site. Instead, it is frequent in literature to adopt the notation (A&C) for the sites occupied by the X atoms, (B) for the site occupied by Y atoms, and (D) for the Z atoms, as shown in Fig. 1(a). Besides ordered structure, HAs are also found to be crystallized in B2 or A2 type disordered phase. B2-disorder happens when the sites of Y and Z atoms become equivalent and forms a different cubic structure with reduced symmetry under space group $Pm\bar{3}m$, as shown in Fig. 1(b). This usually occurs in off-stoichiometric Heusler systems. A complete or A2-type disorder is observed, when all the sites of X, Y and Z atoms become equivalent. In this case, a bcc lattice is formed with reduced symmetry under space group $Im\bar{3}m$, as shown in Fig. 1(c). The ideal $L2_1$ phase can be identified experimentally, when (111) and (200) superlattice peaks are present in the XRD profile. In the case of B2 disordered phase, the (111) reflection is absent in XRD profile. Whereas, for the A2 type fully disordered structure, both the superlattice peaks (111) and (200) disappear in XRD profile [37]. A significant advancement can be seen in HAs due to new experimental approaches [38]. They are extensively studied in the form of thin films and bulk materials [39-41], although, their study and production at the nanoscale still stay challenging [42]. The size and shape modulated nanostructures may preserve an ingenious role in multiple technological areas, as for examples: spintronics, topological insulators and skyrmionic structures [43]. Among all HAs family, $Fe_2$-based HAs have attracted great interest due to their high $M_s$ and $T_c$ [13]. The structural and electronic properties of $Fe_2YAl$ type HAs have been intensively investigated [44-49] but their studies are very limited in nano regime, especially in $Fe_2CoAl$. Recently, thermoelectric property of $Fe_2CoAl$ ribbon sample was explored [50], It exhibited a high $M_s$ of 135 emu/g and $T_c$ of around 1000 K along with a negative seebeck coefficient of -20 µV/K. In this article, we study the structural, magnetic and MC properties of $Fe_2CoAl$ (FCA) nanoalloy. The powder X-ray diffraction (XRD) study suggests that FCA nanoalloy is crystallized in A2 disordered single phase, and the observed lattice constant is to be 5.74 Å. The morphological and microstructural studies reveal that the NPs are highly crystalline in nature. A relatively large value of $M_s$ and $T_c$ are observed in the present system; moreover, the $M_s$ value is very stable even up to room temperature, which is desired for spintronics applications. We further systematically explore its magnetic phase transition (MPT) and MC property across the phase transition ($T_c$). A large peak value of the magnetic entropy change $(-\Delta S_M)_{peak}$ together with the high refrigeration capacity (RCP) and a broad working temperature range make FCA nanoalloy a potential candidate not only for spintronics but also for multistage magnetic refrigeration.

## 2. Experimental details

$Fe_2CoAl$ Heusler alloy nanoparticles were synthesized using co-precipitation method as described in ref. [51] with slight modification. We have used $Fe(NO_3)_3 \cdot 9H_2O$, $CoCl_2 \cdot 6H_2O$, and $Al_2(NO_3) \cdot 18H_2O$ as a precursor salts, which were directly purchased from the Sigma-Aldrich. In a typical preparation of the FCA-nanoalloy, all the precursors with an appropriate molar ratio had dissolved in 50 ml CH₃OH and dried for 10 hours at 100 °C. Thereafter, the dried powder was placed inside a tubular furnace and heated up to 850 °C for the 5 hours, in presence of $H_2$ environment. The crystalline phase at the room temperature was identified by high-resolution (HR) X-ray diffraction (XRD) technique using CuKα radiation (λ = 1.542 Å). The XRD pattern in 2θ interval from 10-100° was recorded with a step size of 0.02°. The microstructural studies were performed using field-emission scanning electron microscope (MERLIN), and HR-transmission electron microscope (TEM). The purity and composition of the elements were determined from the energy dispersive X-ray analysis (EDAX). To study the magnetic and magnetocaloric properties of FCA-NPs, the physical property measurement system (PPMS) of Cryogenics, UK and the vibrating sample magnetometer (VSM) of Lake Shore, USA with working temperature range of 5-300 K and 300-1273 K respectively, were used. In high-temperature magnetic measurements, the powder sample was placed into a disposable boron nitride (BN) cup as provided by Lake Shore, USA with the system. The isothermal magnetization vs magnetic field data were recorded for several temperatures near the magnetic phase transition temperature ($T_c$). Prior to the T dependent magnetization M(T) measurements, the sample was primarily cooled to 300 K from 1000 K in a zero field, which is called zero field cooled (ZFC) condition. Subsequently, data were recorded with increasing T up to 1000 K in the presence of applied field of 100 Oe.

## 3. Results and discussion

### 3.1. Microstructural, morphological and compositional analysis of Fe₂CoAl nanoalloy

The experimental and simulated X-ray diffraction patterns of FCA-NPs are shown in Fig. 2. The absence of (111) and (200) reflections in experimental curve revealed that $L2_1$ phase was not formed in FCA-NPs. The Bragg peaks (220), (400), (422) and (440), observed at 2θ = 44.68°, 65.14°, 82.47°, and 99.29°, respectively, conclude that the sample was crystallized in A2-disordered phase [52]. The calculated lattice constant from the main peak of the XRD profile was found to be $a$ = 5.739 ± 0.0084 Å. This closely matches with the theoretical value of the bulk $Fe_2CoAl$ [13]. It can clearly be seen from the FESEM micrograph (Fig. 3(a)) of FCA-NPs that the particles are densely agglomerated in the form of big clusters, indicating highly magnetic nature of the NPs. To check the purity and composition of the sample, we have done the EDAX analysis on NPs, as shown in Fig. 3(b). This confirms that the particles are nearly stoichiometric and pure in phase. The extra peaks are from the Si-substrate and gold coating, used at the time of sample preparation. The HRTEM image shown in Fig. 4(a) indicates that the particles are spherical in shape. The image is analyzed for particle size distribution using the ImageJ software. The histogram of size-distribution, along with the fitted Gaussian profile, is shown in Fig. 4(b).

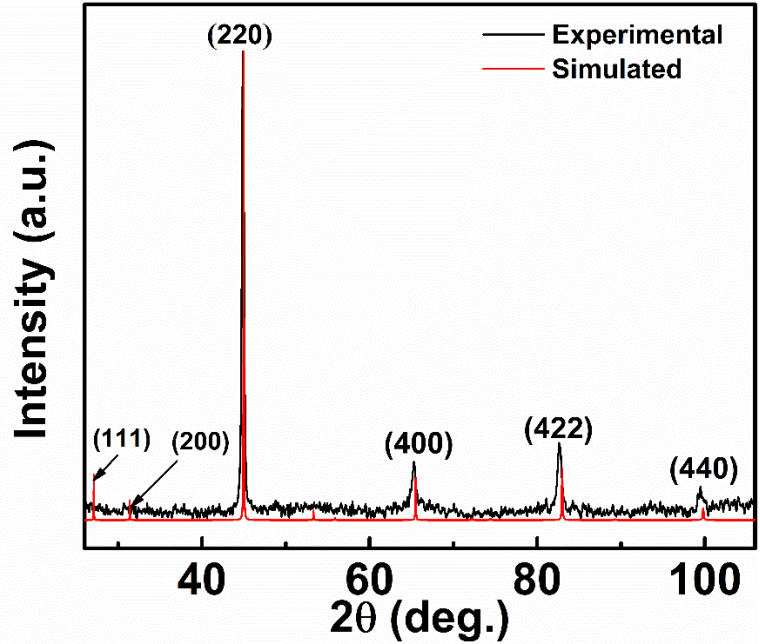

**Fig. 2.** The experimental and simulated x-ray diffraction pattern of FCA-NPs annealed at 850 °C for five hrs.

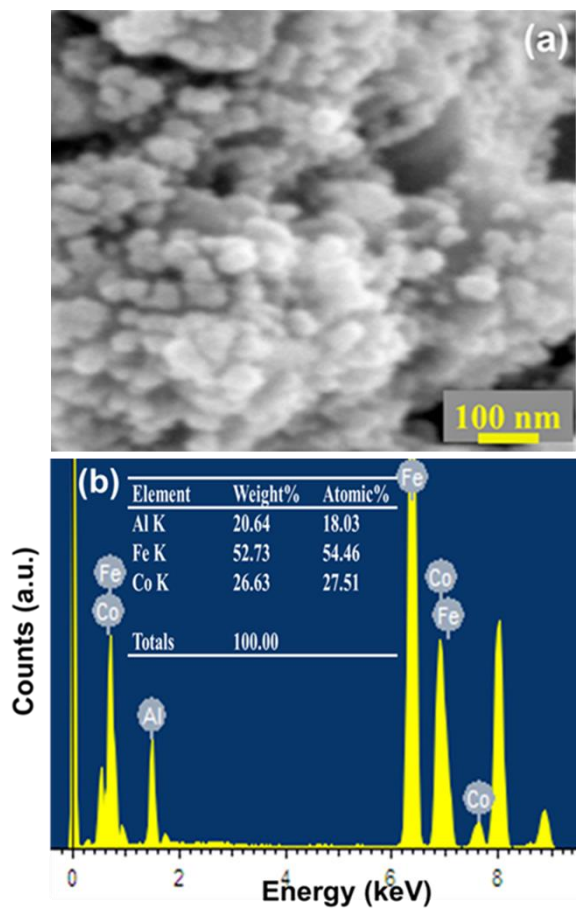

**Fig. 3.** (a) FESEM image of the FCA-NPs annealed at 850 °C for five hours and (b) EDAX spectrum; the inset depicts elemental composition of Fe$_2$CoAl nanoparticles.

According to the histogram, the NPs are of size 14 ± 7 nm. The selected area electron diffraction (SAED) pattern (Fig. 4(c)) encompassed concentric rings with dots confirms that the particles are crystalline. The first four rings are indexed with the (hkl) values (220), (400), (422) and (440), and hence are consistent with the XRD results. The image of a single particle with an even higher resolution have been shown in Fig. 4(d). The evidently visible lattice fringes confirm the high crystallinity of the particles. An enlarged portion of the image is shown in the inset of Fig. 4(d). The interplanar spacing of 2.07 Å, as shown, is equivalent to the (220) plane of the cubic Heusler phase of Fe$_2$CoAl.

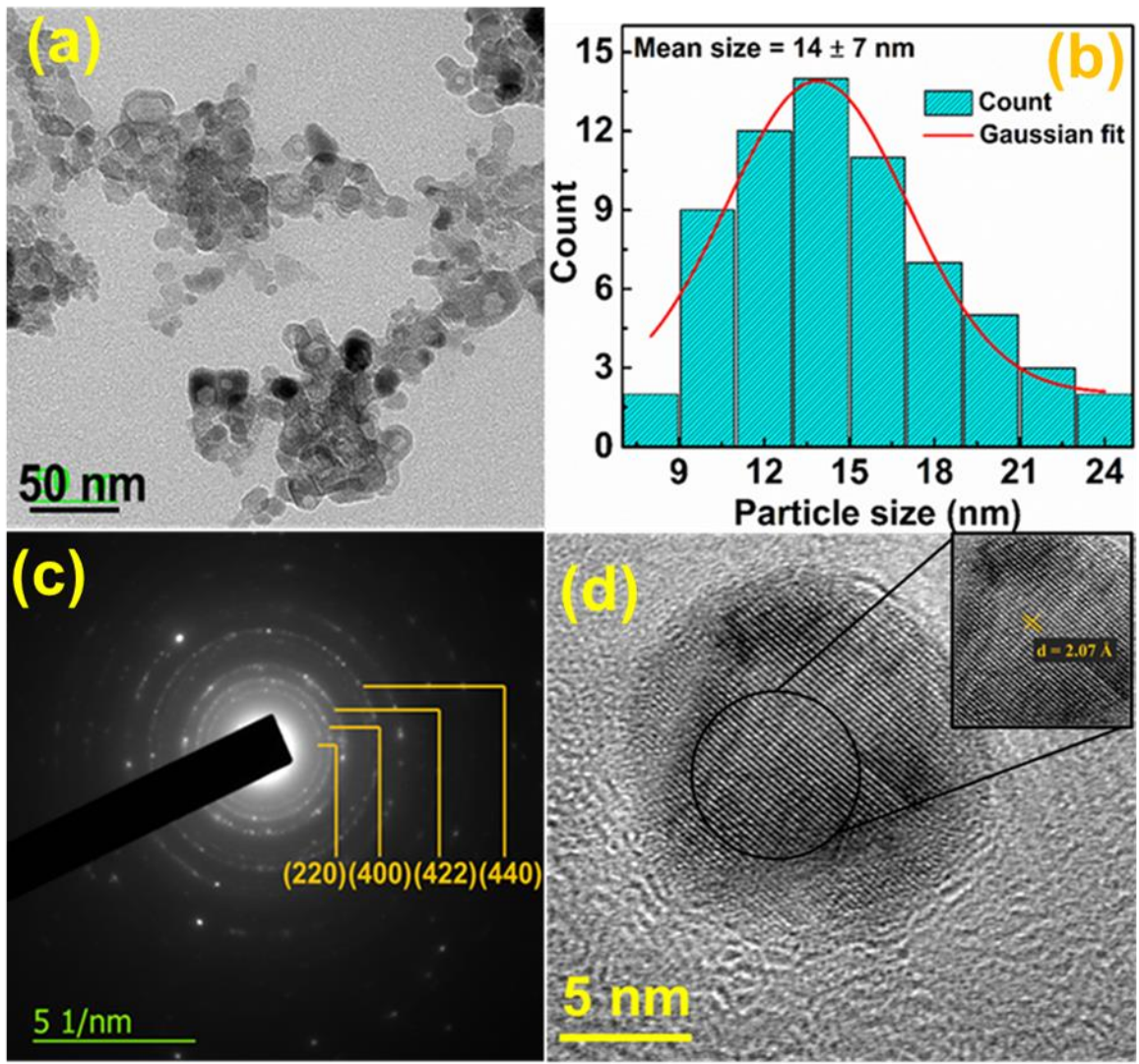

**Fig. 4.** (a) Transmission electron microscopy (TEM) image of the Fe$_2$CoAl nanoalloy. (b) Histogram of particle size. (c) SAED pattern with indexing of initial four rings. (d) High-resolution image displaying the lattice planes and crystallinity.

### 3.2. Magnetic behavior of Fe$_2$CoAl Heusler nanoalloy

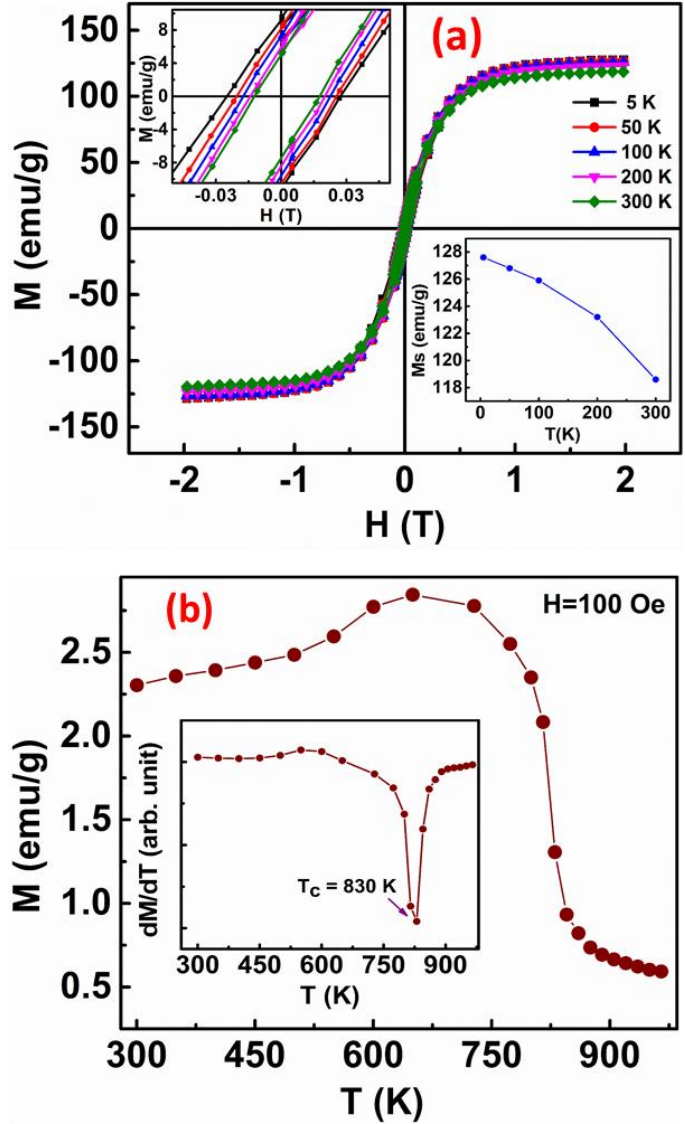

**Fig. 5.** (a-b) Temperature and field dependent magnetization curves of FCA-NPs; the upper and lower insets of figure 5(a) show the magnified version near the zero field region and the variation of saturation magnetization with temperature, respectively. Inset of figure 5(b) shows the dM/dT vs T curve.

The field-dependent magnetization M (H) curves at different temperatures starting from 5 to 300 K are shown in Fig. 5(a). The low-temperature (5 K) $M_s$ was found to be 127.6 emu/g and converted into 4.5 $\mu_B$/f.u. This is larger than the Slater Pauling (SP) value of 4 $\mu_B$/f.u. [53], and slightly less than the bulk value as reported by V. Jain et al. [54]. The reason of deviation from the SP value might be due to the disorder present in this nanometric sample. As evident from the inset of M (H) curve, the finite value of the coercivity ($H_c$) and the remanence ($M_r$) is indicating ferromagnetic behavior of the NPs. A comparison of all the values of $M_s$, $M_r$, and $H_c$ is tabulated in Table 1. The T dependence of the $M_s$ curve, the lower inset of Fig. 5(a), reveals that $M_s$ is stable even up to room temperature, which is good for spintronics application. The temperature dependent magnetization, M (T) curve at the applied field of 100 Oe, under zero field-cooled condition (ZFC) is shown in Fig. 5(b). M is increased with respect to the T and reached maximum at 650 K. The ZFC curve of FCA-NPs was smooth and show a clear FM to PM phase transition at the $T_c$. We have also measured the M (T) curve under field cooled (FC) condition (not shown here) and found no significant thermal hysteresis in ZFC-FC curves of FCA-NPs, which suggests that the phase transition is of second-order [55]. The $T_c$ was calculated from the first derivative of M (T) versus T curve (see inset of figure 5(b)), and was found to be 830 K.

### 3.3. Magnetocaloric properties of Fe$_2$CoAl Heusler nanoalloy

To analyze the MC properties of FCA-NPs, isothermal magnetization curves in field increasing (0 Oe-20 kOe) mode, were recorded at various temperatures from 795-851 K with a temperature interval of 4 K, across the $T_c$, as shown in Fig. 6(a). The increasing nature of the magnetization up to 5 kOe subsequently becomes constant, and then decreases with respect to the temperature increment indicating a magnetic transition from FM to PM phase. Moreover, a huge change of the magnetization in the temperature range of 827-831 K suggests a large MC effect in FCA-NPs. Here we did not find any significant magnetic hysteresis with field decreasing (20 kOe - 0 Oe) mode (not shown here). This reversible nature of magnetic phase transition is vital for magnetic refrigeration. Arrott plot or $M^2$ versus H/M curve [56] was also constructed, to check if the Landau mean field theory for the MPT is suitable for FCA-NPs, as shown in Fig. 6(b).

| Temperature (K) | Sat. magn. $M_s$ (emu/g) | Remanence $M_r$ (emu/g) | Coercivity $H_c$ (Oe) | Curie temperature (K) |
|---|---|---|---|---|
| 5 | 127.6 | 9.2 | 259 | 830 [this work] |
| 50 | 126.8 | 8.2 | 229 | 990 [50] |
| 100 | 125.9 | 7.3 | 200 | |
| 200 | 123.2 | 5.9 | 171 | |
| 300 | 118.6 | 5.0 | 150 | |
| **Bulk Sat. magn.** | | | 4.7 $\mu_B$/f.u. [54] | |
| **Slater Pauling $M_s$ value** | | | 4.0 $\mu_B$/f.u. [53] | |

**Table 1.** Temperature dependence of $M_s$, $M_r$, $H_c$, and $T_c$ of FCA-NPs; other theoretical and experimental values are also tabulated for a comparison.

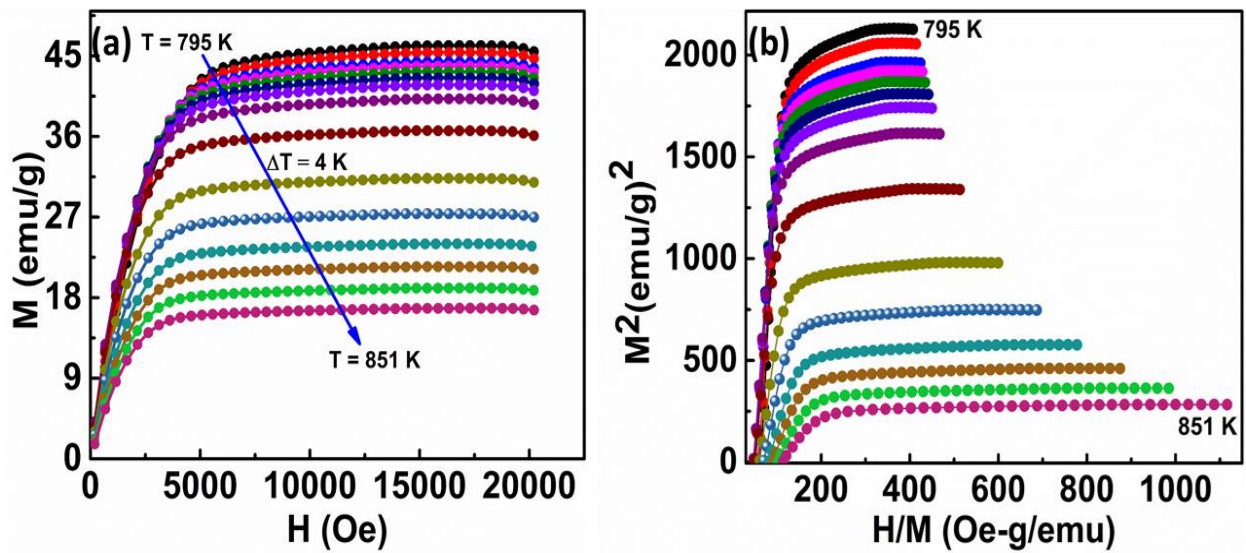

**Fig. 6**. (a) M (H) curves of Fe$_2$CoAl Heusler alloy nanoparticles measured around the phase transition and (b) Arrott curve of the isotherms near the $T_c$.

It is widely accepted that the conventional Arrott curves would exhibit parallel straight lines for the critical exponents of $\gamma = 1$ and $\beta = 0.5$, and the critical temperature ($T_c$) will be defined exactly at the line which passes through the origin. The downward curvature along with the nonlinear behavior even at high-field region was observed. Apparently, we conclude that the mean field theory of phase transition is not valid in present system and a complex behavior of the critical exponents near the $T_c$ is expected. Moreover, the positive slope of Arrott plot following Banerjee's criterion [57] indicates FM to PM phase transition is of second order. The change in magnetic entropy ($-\Delta S_M$) was calculated using Maxwell equation [58]:

$$\Delta S_M = S(T,H) - S(T,0) = \int_0^H \left(\frac{\partial M}{\partial T}\right)_H dH \quad (1)$$

The T dependence of magnetic entropy change ($-\Delta S_M$) is presented in Fig. 7. A positive anomaly in $-\Delta S_M$ vs T curve was observed at 830 K, which is much closer to its $T_c$. The maxima in the change in magnetic entropy vs temperature curve at a magnetic field of 20 kOe corresponds to about 2.65 J/Kg-K. It is clear from the inset of Fig. 7 that the peak value of the magnetic entropy change, $(-\Delta S_M)_{peak}$ increases linearly with temperature; therefore, a large value of $-\Delta S_M$ is expected for higher fields but we could not measure it due to the temperature limitation of our system. The usefulness of magnetic refrigerants can be evaluated by a parameter: relative cooling power (RCP), which measures the amount of heat transferred between hot and cold reservoirs and is defined as RCP = $-$ ($\Delta S_M)_{peak}$ x $\Delta T_{FWHM}$, where $\Delta T_{FWHM}$ denotes the full-width at half-maximum of the change in $-\Delta S_M$ vs T curve [59]. $\Delta T_{FWHM}$ also represents the span of working-temperature in the system, which was found to be around 17 K; such broad working temperature range is highly desired for magnetic refrigeration. As Engelbrecht *et al.* [60] previously suggested that a material having a broad peak of $-\Delta S_M$ is much better than that of materials with a sharp peak of magnetic entropy change for cooling applications and therefore, such materials with broad temperature distribution, are much attractive for MC application. The observed value of RCP is as large as 44 J/Kg at the magnetic field change of 20 kOe, suggesting a large heat conversion in magnetic refrigeration cycle. As observed from the inset of Fig. 7, and Fig. 8, both the change in magnetic entropy and RCP values have shown a linear dependence with the applied field in the range of 0-20 kOe. This increasing nature of $(-\Delta S_M)_{peak}$ and RCP with $\Delta H$ along with the broader working temperature range would be suitable for Ericsson-cycle refrigeration application [61]. Thus exploitation of magnetic and magnetocaloric properties in $Fe_2$-based Heusler nanoalloys may be useful for not only spintronics but also for multistage magnetic refrigeration technology [34, 62]. A comparison of these values with previous magnetic refrigerants is presented in Table 2.

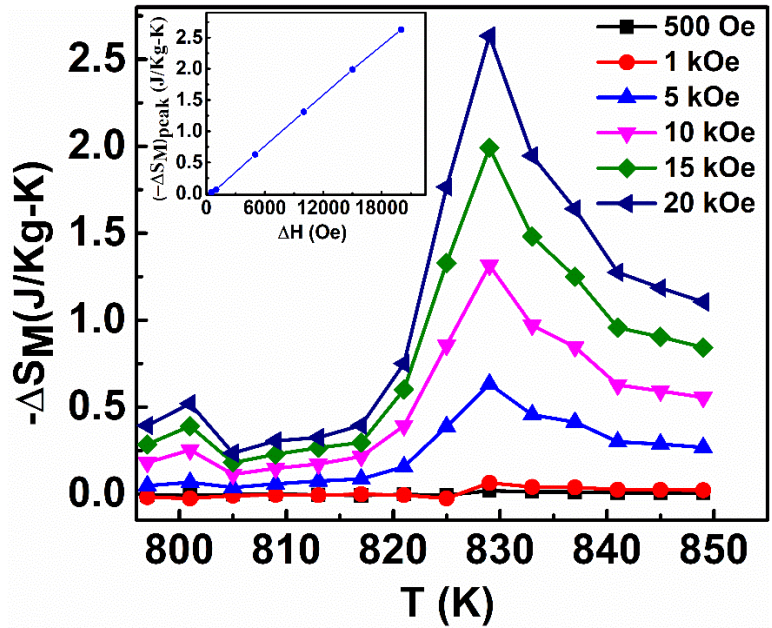

**Fig. 7.** $-\Delta S_M$ versus temperature curve for $Fe_2CoAl$ Heusler nanoalloy; The inset shows the change of $(-\Delta S_M)_{peak}$ value with respect to the changing field ($\Delta H$).

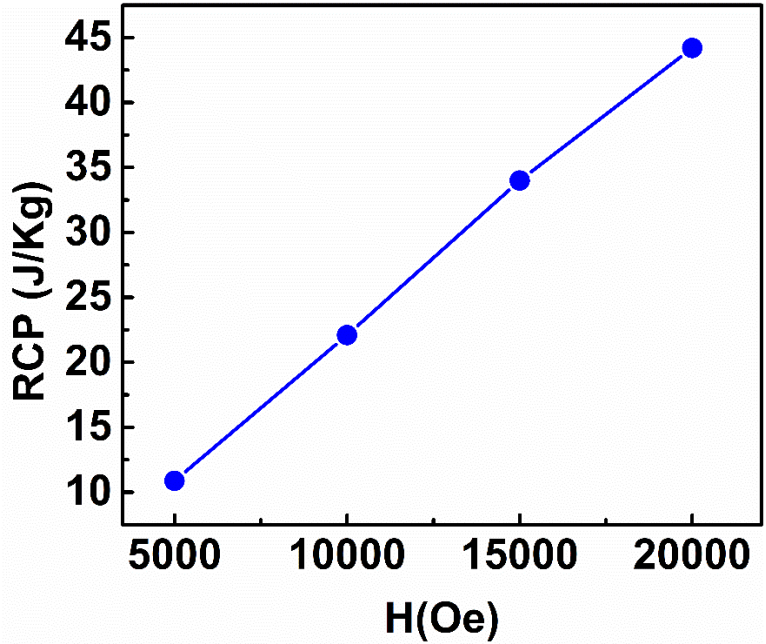

**Fig. 8.** Field dependence of the relative cooling power (RCP) across the phase transition ($T_c$)

| Materials | | $-\Delta S_M$ (J/Kg-K) | Working temperature span, $\Delta T_{FWHM}$ (K) | RCP (J/Kg) | Field range ($\Delta H$) in kOe | $T_c$ (K) | Nature of the Phase transition | Refs. |
|---|---|---|---|---|---|---|---|---|
| **Heusler nanoparticles** | $Fe_2CoAl$ | 2.65 | 17 | 44 | 20 | 830 | Second-order | present work |
| | $Co_2FeAl$ | 15 | 6 | 89 | 14 | 1261 | Second-order | [17] |
| **Bulk Heusler alloys** | $Mn_{1-x}Cr_xCoGe$ | ~28.5 | – | – | 50 | 322 | First-order | [63] |
| | $MnCo_{1-x}Zr_xGe$ | 7.2 | – | 266 | 50 | 274 | First-order | [64] |
| | $Mn_{1-x}Al_xCoGe$ | 12 | – | 303 | 50 | 286 | First-order | [65] |
| | $Co_2Cr_{0.25}Mn_{0.75}Al$ | 3.5 | – | 285 | 90 | 720 | Second-order | [34] |

**Table 2.** A comparison of the $(-\Delta S_M)_{peak}$, RCP, $\Delta T_{FWHM}$, $\Delta H$, $T_c$, and nature of the phase transition in Fe$_2$CoAl nanoalloy and other MC materials.

In Co$_2$FeAl Heusler NPs [66], we have observed a giant magnetocaloric effect around 1252 K. However, such a high T$_c$ and sharp peak in $(-\Delta S_M)_{peak}$ vs temperature curve make it less efficient for magnetic refrigeration. This further opens a way of hunting the new Heusler compounds, especially, in nano regime. Our $(-\Delta S_M)_{peak}$, value is higher than that of the Co$_2$Cr$_{0.25}$Mn$_{0.75}$Al ( Table 2), and RCP value is at least comparable on the same field scale. The present values are also considerably larger than that of the other Co$_2$-based full Heusler alloys [67]. On the contrary, a huge RCP value of 400 J/Kg at 50-kOe magnetic field was reported in Gadolinium [68]. Nevertheless, it is not suitable for commercial purposes due to its high cost, and therefore Ni-based Heusler compounds [69-72] are extensively investigated for better performing magnetocaloric effect in recent years. Though, most of them exhibit a first-order phase transition causing a thermal and magnetic hysteresis near the T$_c$. The present MC study suggests that the Fe$_2$CoAl nanoalloy might be suitable for multistage magnetic refrigeration technology and further opens a window of research on how its magnetic and magnetocaloric properties can be tuned by size and shape modulation of the nanoparticles, which we aim in our future work. A master curve or universal curve was proposed by Franco et al. [73] for the MC materials exhibiting second-order phase transition near the T$_c$. The curve was defined as the normalized entropy change $\Delta S'_M$ with respect to rescaled temperature ($\theta$) where $\Delta S'_M = \Delta S_M / \Delta S_M^{peak}$ and $\theta$ defined as:

$$\theta = \left\{ \frac{-(T-T_c)}{(T_{r1}-T_c)}, \text{for } T \leq T_C; \frac{(T-T_c)}{(T_{r2}-T_c)}, \text{ for } T > T_C \right\} \quad (2)$$

In above equation, Curie temperature (T$_c$) denotes the temperature at the peak value of the magnetic entropy change ($\Delta S_M^{peak}$). Reference temperatures $T_{r1}$ and $T_{r2}$ were taken as the temperatures analogous to $\frac{1}{2}\Delta S_M^{peak}$ below and above the Curie temperature (T$_C$). As seen from Fig. 9, $\Delta S'_M$ vs $\theta$ curves of FCA-NPs, taken at different applied fields, have been merged into a single universal curve, which confirms a second order magnetic phase transition across the T$_c$.

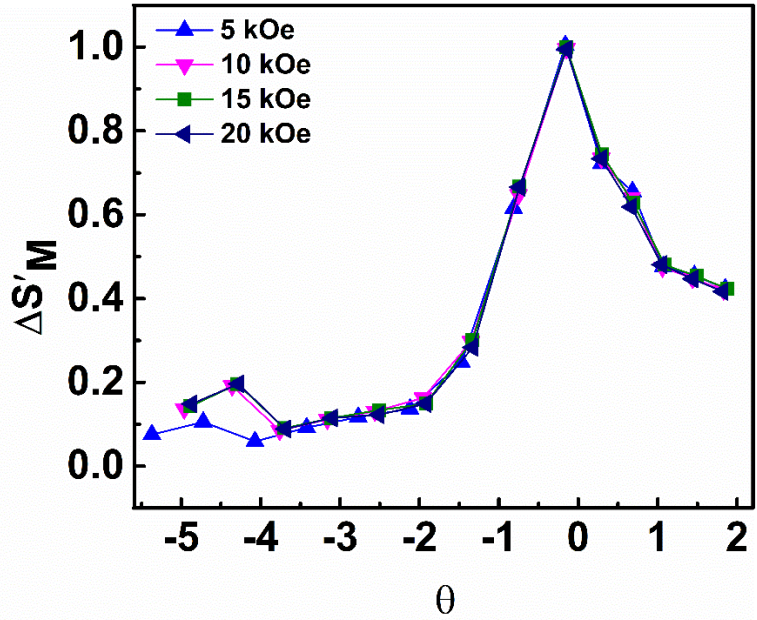

**Fig. 9.** The magnetic entropy ($\Delta S'_M$) as a function of rescaled temperature (θ) near $T_c$ for Fe$_2$CoAl Heusler nanoalloy.

Such behavior of phenomenological curves was observed previously in second-order phase transition materials [66, 73, 74]. This study recommends of using this large MC material with a broad working temperature range in magnetic refrigeration, as in multi-stage magnetic refrigeration where cooling from high temperature is desired in more than one stages [75]. Such multi-stage magnetic refrigerators are highly appropriate for industrial application, where the low temperature of one stage acts as upper temperature for the next stage [26, 75].

## 4. Conclusions

Fe$_2$CoAl Heusler nanoparticles of size 14 nm were grown using co-precipitation method. The microstructure of the NPs was examined by XRD, FESEM, HRTEM, and SAED. Structural characterizations confirmed the crystalline nature of particles with A2-disorder, and uniform chemical distribution of the elements. The magnetic characterization on the FCA-NPs supports their soft ferromagnetic character with high $M_s$ of about 127.6 emu/g (or 4.5 $\mu_B$/f.u.) and high $T_c$ of 830 K. The peak value of the change in magnetic entropy ($-\Delta S_M$) vs temperature curve at a magnetic field of 20 kOe corresponds to about 2.65 J/Kg-K, and the observed value of refrigeration capacity (RCP) is as large as 44 J/Kg. To analyze the MPT, a detailed analysis of magnetization is performed. The Arrott plot and the nature of the universal curve accomplish that the FM-PM phase transition is of second order.

**Notes:** The authors declare no competing financial interest.

**Acknowledgements:**

Aquil Ahmad acknowledges the University Grants Commission, New Delhi and Ministry of Education (MoE), India for providing the research fellowship. A. Ahmad also acknowledges the Central Research Facility (CRF), and the Departmental lab facility of IIT Kharagpur, India for the sample characterizations and DFT computations. A K Das acknowledges the financial support from the Department of Science and Technology (DST), India (Project No. EMR/2014/001026).